\documentstyle[12pt]{article}

\begin{document}

\newcommand{\beq}{\begin{equation}}
\newcommand{\eeq}{\end{equation}}
\newcommand{\beqa}{\begin{eqnarray}}
\newcommand{\eeqa}{\end{eqnarray}}
\newcommand{\beqar}{\begin{eqnarray*}}
\newcommand{\eeqar}{\end{eqnarray*}}
\newcommand{\tr}{{\rm tr}}
\newcommand{\be}{\beta}
\newcommand{\al}{\alpha}
\newcommand{\ie}{{\it i.e.,}\ }
\newcommand{\eg}{{\it e.g.,}\ }
\newcommand{\ch}{{\cal H}}
\newcommand{\ssc}{\scriptscriptstyle}

\def\d{\delta}
\def\eps{\epsilon}
\def\heps{\hat{\epsilon}}
\def\beps{\bar{\epsilon}}
\def\rL{\widetilde{L}}
\def\tc{\tilde{\chi}^a}
\def\tchi{\tilde{\chi}}
\def\phis{\phi}    
\def\phim{\psi}    
\def\sL{\widetilde{L}} 
\def\fL{L}         

\begin{titlepage}
\thispagestyle{empty}
\begin{flushright} {\small McGill/93--22;} {\small NSF--ITP--93--152;}
                   {\small UMDGR--94--75}\\
                   {\small gr--qc/9312023}\\
\end{flushright}

\bigskip

\begin{center}
{\bf \huge On Black Hole Entropy}\\[2ex]

\bigskip

Ted Jacobson$^{a,b,}$\footnote{jacobson@umdhep.umd.edu},
Gungwon Kang$^{a,b,}$\footnote{eunjoo@wam.umd.edu},
and Robert C. Myers$^{a,c,}$\footnote{rcm@hep.physics.mcgill.ca}\\
\medskip

$^a${\small Institute for Theoretical Physics, University of California,
Santa Barbara, CA 93106}\\

$^b${\small Department of Physics, University
of Maryland, College Park, MD 20742--4111}\\

$^c${\small Department of Physics, McGill University, Montr\'{e}al, Qu\'{e}bec,
Canada H3A 2T8}\\

\vfill

\begin{abstract}
{\tenrm\baselineskip=12pt
 \noindent
Two techniques for computing black hole entropy
in generally covariant gravity theories including
arbitrary higher derivative interactions are studied.
The techniques are Wald's Noether charge
approach introduced recently, and a field redefinition
method developed in this paper. Wald's results are extended
by establishing that his local geometric expression for
the black hole entropy gives the same result when evaluated
on an arbitrary cross-section of a Killing horizon (rather
than just the bifurcation surface). Further, we show that
his expression for the entropy is not affected by ambiguities
which arise in the Noether construction.
Using the Noether charge expression,
the entropy is evaluated explicitly for black holes in
a wide class of generally covariant theories.
For a Lagrangian of the functional form
$\sL=\sL(\phim_m,\nabla_a\phim_m,g_{ab},R_{abcd}, \nabla_eR_{abcd})$,
it is found that the entropy is given by
$$
S=-2\pi\oint \left(Y^{abcd}-\nabla_eZ^{e:abcd}\right)
\heps_{ab}\heps_{cd}\beps
$$
where the integral is over an arbitrary cross-section
of the Killing horizon, $\heps_{ab}$ is the binormal to the
cross-section, $Y^{abcd}=\partial \sL/\partial R_{abcd}$,
and $Z^{e:abcd}=\partial \sL/\partial \nabla_eR_{abcd}$.

Further, it is shown that
the Killing horizon and surface gravity of a stationary black
hole metric
are invariant under field redefinitions of the metric of the form
$\bar{g}_{ab}\equiv g_{ab} + \Delta_{ab}$, where $\Delta_{ab}$ is
a tensor field constructed out of stationary fields.
Using this result, a technique is developed
for evaluating
the black hole entropy in a given theory in terms of that of another
theory related by field redefinitions. Remarkably, it is established
that
certain perturbative, first order, results obtained with this method
are in fact {\it exact}.
A particular result established in this fashion is that a
scalar matter term of the form $\nabla^{2p}\phis\,\nabla^{2q} \phis$
in the Lagrangian makes no contribution to the black hole entropy.

The possible significance of these results for the problem of
finding the statistical origin of black hole entropy is discussed.}

\end{abstract}
\end{center}

\vfill

\end{titlepage}
\pagebreak

\section{Introduction}

Black hole thermodynamics seems to hint at some profound
insights into the character of gravity in general, and
quantum gravity in particular. The hope is that further
study will reveal something about the nature of quantum
gravity. One direction to pursue is to investigate the
stability of black hole thermodynamics under perturbations
of the classical (Einstein) theory that are expected on
general grounds as a result of quantum effects.

Whatever the ultimate nature of quantum gravity, there
should be an effective Lagrangian that describes the
dynamics of a classical ``background field" for sufficiently
weak fields at sufficiently long distances. Such a low
energy effective action will presumably be generally
covariant, and will have higher curvature terms, and also
higher derivative terms in the metric and all other matter
fields. For example, such interactions naturally arise from
renormalization in the context of quantum field
theory\cite{field}, and in the construction of an effective
action for string theory\cite{string}. While such actions
are pathological if taken as fundamental, they can define
benign perturbative corrections to Einstein gravity with
ordinary matter actions. Let us also add at this point that
many of the recent candidates for a theory of quantum
gravity, especially those which attempt to unify gravity
with other interactions, are theories in higher dimensional
spacetimes. Thus in the following investigation, we will
allow spacetime to have an arbitrary dimension, $D$.

The question we would like to address is whether
the laws of black hole thermodynamics are consistent with
all such effective actions, or whether perhaps consistency
with these laws picks out a preferred class of potential
``corrections" to classical gravity.
Most recent efforts have been devoted to calculating $S$
for explicit black hole solutions in various theories.
It is well known that in general
the standard relation,
$S=A/(4G)$, of Einstein gravity no longer
applies\cite{well}.
Until recently though, it
had not even been established (except for special
cases, {\it e.g.,} Lovelock gravity\cite{love}) that the
entropy can be expressed as a local functional evaluated at
the horizon. Now several researchers have shown that $S$ is
indeed local in general, and have provided various
techniques to compute it\cite{wald1,visser,SuWald}. In this
paper, we will compute $S$ for a wide class of theories,
using both Wald's Noether charge technique\cite{wald1}, and
a method exploiting field redefinitions that we developed
prior to the recent appearance of more powerful and general
techniques\cite{wald1,visser}. In addition, we will extend
the results of \cite{wald1} by showing why ambiguities in
the Noether charge construction
do not affect the entropy, and
establishing an expression for the entropy which is valid on
{\it arbitrary} cross-sections of the horizon (rather than
just on the bifurcation surface).

A key concept in what follows is the notion of a Killing
horizon. A Killing horizon is a null hypersurface whose null
generators are orbits of a Killing field. In four
dimensional Einstein gravity, Hawking proved that the event
horizon of a stationary black hole is a Killing
horizon\cite{killhorz}. This proof can not obviously be
extended to higher curvature theories (however no counter-examples
are known, including some non-static
solutions\cite{campbell}).
If the horizon generators of a Killing horizon are geodesically
complete to the past (and if the surface gravity is nonvanishing),
then the Killing horizon
contains a ($D$--2)-dimensional spacelike cross section $B$
on which the Killing field $\chi^a$
vanishes\cite{raczwald}.
$B$ is called the {\it bifurcation surface}.
Such a bifurcation surface is fixed under the Killing flow,
and lies at the
intersection of the two null hypersurfaces that comprise the
full Killing horizon. (For example, in the maximally extended
Schwarzschild black hole
spacetime, the bifurcation surface is the 2-sphere of area $16\pi M^2$
at the origin of Kruskal $U$-$V$ coordinates.)
The techniques employed for computing black hole entropy
in this paper all apply only to black holes with
bifurcate Killing horizons.
For a spacetime containing an
asymptotically stationary black hole
that forms by a collapse process
there is certainly no bifurcation surface. However,
the operative assumption is then that
the stationary black hole which is
asymptotically approached can be extended to the past
to a spacetime with a bifurcate Killing horizon.
Investigation of the
validity of this fundamental assumption is important but
will be left to other work.

The zeroth law of black hole thermodynamics
states that for a stationary
event horizon the surface gravity $\kappa$, which is
proportional to the black hole's temperature (in any theory
of gravity), is constant over the entire horizon. If
$\chi^a$ is the null generator of the horizon, $\kappa$ is
defined by $\chi^b\nabla_b\chi^a =\kappa \chi^a$.
This constancy of $\kappa$
on Killing horizons
has been proven for
Einstein gravity\cite{barcarhaw}
with matter satisfying the dominant energy condition, but this proof
does not readily extend to the present context of
higher curvature theories.
If one {\it assumes} (as we do) a bifurcate Killing horizon however,
then
constancy of the surface gravity is easily seen to hold independently
of any field equations. Conversely, if the surface gravity
is constant and nonvanishing on a Killing horizon, then
the horizon can be extended to a bifurcate horizon \cite{raczwald}.

The first law of black hole thermodynamics takes the form
\beq
\frac{\kappa}{2\pi}\delta S=\delta M-\Omega^{(\al)}\,\delta J_{(\al)}
-\cdots\ \ .
\label{first}
\eeq
$M$, $J_{(\al)}$ and $\Omega^{(\al)}$ are the black hole
mass, angular momentum, and the angular velocity of the
horizon\cite{foot1}. The ellipsis indicates possible
contributions from variations of other extensive parameters
which characterize the black hole (\eg electric or magnetic
charge). For Einstein gravity, the entropy $S$ is
one-quarter the surface area of the horizon,
$S=A/(4G)$. Eq.~(\ref{first}) then has the rather remarkable
feature that it relates variations in properties of the
black hole as measured at asymptotic infinity to a variation
of a geometric property of the horizon. Given the recent
results of ref.'s \cite{wald1,visser}, one now knows that
although the precise expression for the entropy is altered,
it remains a quantity localized at the horizon for arbitrary
theories of gravity, and so this aspect of the first law is
preserved.

The remainder of the paper is organized as follows: In
sect.~2, we describe Wald's result that the entropy is a
Noether charge\cite{wald1}, generalize it to arbitrary
horizon cross-sections, and
discuss the extension to nonstationary black holes.
We also apply this technique to compute $S$ for a
certain wide
class of Lagrangians.
Sect.~3 introduces the field redefinition
technique, and illustrates it with examples. Sect.~4
presents a discussion of our results
and their possible implications for
the problem of the statistical origin of black hole entropy.
Throughout the paper, we consider only asymptotically
flat spaces, and we employ the conventions of \cite{exam}.

\section{Entropy as a Noether charge}
Wald \cite{wald1} has recently derived a general  formula
for the entropy of a stationary black hole based on a
Lagrangian derivation of the first law of black hole
mechanics. The results of \cite{wald1} apply to black  holes
with bifurcate Killing horizons in any diffeomorphism
invariant theory in any spacetime dimension. (The entire discussion
of this section refers to stationary black holes of this type,
except in sect.~2.3.)
Wald finds that that the entropy is given by
\begin{equation}
S=2\pi\oint Q\ ,
\label{S}
\end{equation}
where the integral is over the bifurcation surface of the
horizon. The $(D-2)$-form  Q is the ``Noether potential"
(defined below) associated with
the Killing field $\tilde{\chi}^a$
that is null on the horizon and normalized to have unit
surface gravity. As we shall show below, one obtains the
same result  for the entropy if $Q$ is integrated over {\it
any} cross-section of  the horizon.

The entropy (\ref{S}) is not expressed
in terms of {\it only} the  dynamical fields and their
derivatives, since by construction $Q$ involves the Killing
field $\tilde{\chi}^a$ and its
derivatives. (The  Killing field is of
course determined by the metric, but by an integral
operation rather than differential ones.)
However, as noted in \cite{wald1}, the
explicit dependence on
$\tilde{\chi}^a$ can be eliminated as follows.  First, one
uses the identity $\nabla_c \nabla_a \tilde{\chi}_b = -
R_{abcd} \tilde{\chi}^d$ (which holds for any Killing
vector) to eliminate any second or higher derivatives of
$\tilde{\chi}^a$, leaving only $\tilde{\chi}^a$ and
$\nabla_a\tilde{\chi}_b$.
The term linear in $\tc$ contributes nothing
at the bifurcation surface,
since $\tc$ vanishes there. Moreover, at the bifurcation surface
one has $\nabla_a\tilde{\chi}_b=\heps_{ab}$, where
$\heps_{ab}$ denotes the binormal to $B$.
This allows one to substitute
$\heps_{ab}$ for $\nabla_a\tilde{\chi}_b$ in the expression for $Q$.
Thus, at least at the bifurcation surface, all explicit reference to
the Killing field can be eliminated from $Q$.
Let us denote by $\tilde{Q}$ the form that is obtained from $Q$
in this fashion. Then the expression
$2\pi\oint\tilde{Q}$ evaluated at the bifurcation surface
correctly displays the entropy
as a ``local'' geometric functional
of the metric, the matter
fields and their derivatives. It will be shown below that
in fact $S=2\pi\oint\tilde{Q}$ on
an arbitrary cross-section  of the Killing horizon.

Wald's construction for the entropy has tremendous advantages
compared with methods previously available. One works with
an arbitrary Lagrangian, and there is no need to find the
corresponding Hamiltonian, as in ref.~\cite{SuWald}. Further
there is no need to identify a preferred surface term in the
action. Adding a total derivative to the Lagrangian does
not affect the entropy (\ref{S}), as will be shown below.
Another feature to be noted is that no ``Euclideanization"
is required.

In the remainder of this section, we first sketch the
Noether  charge construction\cite{wallee,wald2}, and explain
why the  attendant ambiguities do not affect the entropy.
Then, we show how the entropy can be expressed as an
integral over an arbitrary cross-section of the horizon,
rather than over the bifurcation surface.
Next, we discuss the possible definitions of entropy for
nonstationary black holes. Finally, we
explicitly compute the black hole entropy for a wide class
of Lagrangians.

\subsection{The Noether potential Q}
\label{Noether}

The $(D$--2)-form $Q$
may be defined as the potential for the corresponding
``Noether current" ($D$--1)-form $J$,
in the case that $J=dQ$. The
symmetry relevant for the black hole entropy is
diffeomorphism invariance. The Noether current associated
with the diffeomorphism generated by a vector field $\xi^a$
is defined as follows\cite{wallee}. Let $\fL$ be a Lagrangian
$D$-form built out of some set of dynamical fields,
including the metric, collectively denoted here by $\phim$.
Under a general field variation $\d \phim$, the Lagrangian
varies as
\begin{equation}
\d \fL=E\cdot \d \phim + d\theta(\d \phim),
\label{dL}
\end{equation}
where ``$\cdot$" denotes a summation over the dynamical
fields including contractions of tensor indices, and $E=0$
are the equations of motion. (The ambiguity
$\theta\rightarrow \theta+d\gamma$ allowed by (\ref{dL}) is
inconsequential -- see below.)

The diffeomorphism invariance of the theory is ensured if,
under field variations induced by diffeomorphisms $\d
\phim={\cal L}_\xi\phim$, one has $\d \fL={\cal L}_\xi
\fL=d\,i_\xi \fL$.\cite{footn}
 The Noether current $J$
associated with a vector field $\xi^a$ is defined by
\begin{equation}
J=\theta({\cal L}_\xi \phim)- i_\xi \fL,
\label{j}
\end{equation}
where $\theta$ is defined by (\ref{dL}). One easily sees
that $dJ=0$, modulo the equations of motion, as a
consequence of the diffeomorphism covariance of the
Lagrangian. Thus, modulo the equations of motion, we have
$J=dQ$, for some $Q$, at least locally. Much more can be
said however, as a consequence of the fact that $J$ is
closed  for {\it all} vector fields $\xi^a$. Namely, there
exists  a unique, globally defined $Q$ satisfying $J=dQ$,
that is a local function of the dynamical fields and a
linear function of $\xi^a$ and its derivatives.\cite{wald2}
Moreover, ref.~\cite{wald2} presents
an inductive algorithm for constructing $Q$ in such a situation.
We call $Q$ the {\it Noether potential}
associated
with $\xi^a$.
The Noether charge for a spacelike ($D$--1)-dimensional
hypersurface $M$
is given by $\int_M J$, and hence in this case,
reduces to the boundary
integral $\int_{\partial M}Q$. Thus the black hole entropy (\ref{S})
is $2\pi$ times the contribution to the  Noether charge
coming from the boundary at the horizon.

There are three
stages at which ambiguity can enter the
above construction of the Noether charge. First, an exact
form $d\alpha$ can be added to the Lagrangian without
changing the equations of motion.
This induces an extra term
${\cal L}_{\xi}\alpha$ in $\theta({\cal L}_\xi \phim)$, and therefore
extra terms $di_{\xi}\alpha$ and $i_{\xi}\alpha$ appear in $J$ and
$Q$, respectively.
Now in the entropy (\ref{S}), $\xi$ is chosen to be the Killing
field $\tc$, and $Q$ is evaluated at the bifurcation surface
where $\tc$=0. Thus the extra term $i_{\tchi}\alpha$ makes no
contribution to the entropy.
In fact the extra term vanishes {\it everywhere} on the horizon --
as will be shown below -- leaving $2\pi\oint Q$ unmodified for
any cross-section of the Killing horizon.

The second ambiguity arises because $\theta$ is defined by
(\ref{dL}) only up to the addition of a closed form $\beta$.
Assuming $\beta$ is closed for {\it all} variations $\delta\phim$
(and that it vanishes when $\delta\phim=0$), then the result of
\cite{wald2} quoted above implies that
$\beta$ has the form $d\gamma$.
Thus $J$ and $Q$ are modified by the addition of
$d\gamma({\cal L}_{\xi}\phim)$ and $\gamma({\cal L}_{\xi}\phim)$,
respectively.
With $\xi$ equal to the Killing field $\tc$,
the extra term $\gamma({\cal L}_{\tilde{\chi}}\phim)$
vanishes because ${\cal L}_{\tilde{\chi}}\phim=0$
for the background fields $\phim$ in a stationary solution.
In this case, it is immediately clear that this ambiguity
will not affect $2\pi\oint Q$ for any slice of the horizon.

The third ambiguity arises because $Q$ is defined by $J=dQ$ only up to the
addition of a closed form $\sigma$. With the same assumptions as for
$\beta$ in the previous paragraph, we similarly conclude that $\sigma$ is
exact. Since the integral of an exact form over a closed surface vanishes,
$Q$ and $Q+\sigma$ yield the same entropy.

\subsection{Arbitrary horizon cross-sections}

There are several reasons why it is important to be able to
evaluate the black hole entropy as an integral over an
arbitrary slice of the Killing horizon rather than only at the
bifurcation surface.  For one thing, if the (approximately
stationary) black hole
formed from collapse, then the bifurcation surface is not
even a part of the  spacetime. As a practical
matter, it may be inconvenient to have to  determine the
geometry and matter fields in the vicinity of the
bifurcation  surface, for instance if the solution is known
in a coordinate system that does not extend all the way to
there. The primary reason though arises because one wants to
have a definition of the entropy that applies to
non-stationary black holes. A clear prerequisite for
such a definition is
that it yield the same result for any cross-section of a
stationary black hole horizon.

In the case of general relativity, Eq.~(\ref{S}) yields one
quarter the area of the bifurcation surface for the entropy.
It is a simple consequence of stationarity
that all cross-sections of the Killing
horizon are isometric, so in particular they have the same
area. Thus  one can deduce that the entropy is one
quarter  the area of {\it any} cross-section of the horizon.
Similarly, in Lovelock  gravity, it was found \cite{love}
that the entropy depends only upon the  intrinsic geometry
of the bifurcation surface. Since all cross-sections  of the
horizon are isometric, the entropy of Lovelock black holes
is given by an intrinsic expression which can be evaluated over any
cross-section  with equal results.

For a general theory, the entropy (\ref{S}) does not  depend
only upon the intrinsic geometry of the horizon, so it is
not  immediately clear what form it will take on arbitrary
cross-sections  of the horizon. Nevertheless,
it is easy to see
that $\oint Q$ is in fact  the same for all cross-sections
of a stationary horizon.
The difference between the
integrals over two cross-sections  is given by the integral
of $dQ=J$ over the segment of the horizon  that is bounded
by them. For a stationary spacetime, eq.~(\ref{j})  yields
$J=-i_{\tilde{\chi}} \fL$, whose pullback to the horizon
vanishes since $\tc$ is tangent to the horizon. Thus the
entropy is indeed given by (\ref{S}) with the integral taken
over {\it any} cross-section of the horizon.

Can the explicit dependence
of the entropy on the Killing field can be eliminated on any slice
of the horizon, as was possible at the bifurcation surface?
Recall that, as explained above, when eliminating the Killing field
from $Q$ to obtain $\tilde{Q}$, Wald used the fact
that $\tc$ vanishes at the bifurcation surface and the fact
that $\nabla_a\tilde{\chi}_b$ is the binormal, neither of
which are true on an arbitrary cross-section. Nevertheless,
we will now show that
although $\tc$
does not vanish the term proportional to $\tc$ vanishes, and
although $\nabla_a\tilde{\chi}_b$ is not  the binormal the
difference between it and the binormal makes no contribution.
That is, the entropy is given by $2\pi\oint \tilde{Q}$ on
{\it any} cross-section of the horizon.

When any higher derivatives of the Killing field are
eliminated by use of the Killing identity, the Noether charge
takes the general form
\begin{equation}
\hat{Q}=B_a \tilde{\chi}^a + C_{ab} \nabla^a \tilde{\chi}^b\, ,
\label{Qgeneral}
\end{equation}
where $B_a$ and $C_{ab}$ are tensor-valued $(D-2)$-forms
that are invariant under the Killing flow.

The pullback to  the horizon of  the form $B_a
\tilde{\chi}^a$ necessarily vanishes everywhere on a
Killing horizon if $B_a$ is regular  at the bifurcation surface. To
see why, note that  the Killing field is tangent to the
horizon and therefore   it defines a flow of the horizon
into itself.  The form $B_a \tilde{\chi}^a$ is invariant
under the Killing flow, so its pullback is an invariant form
on the horizon submanifold.  If it vanishes
at one point on a given horizon generator, it vanishes
everywhere along that generator.
Now it vanishes
on the bifurcation surface  where $\tc$ vanishes, provided
none of the tensor fields out of which  $B_a$ is built are
singular there (regularity at the bifurcation surface
is an implicit assumption already in Wald's
derivation\cite{wald1}).
Furthermore, all horizon generators terminate at the bifurcation
surface.
However, the Killing ``flow" does not flow
anywhere at the bifurcation  surface. Nevertheless, one can
argue by continuity that the  components of the  form
$B_a \tilde{\chi}^a$ are arbitrarily small in good coordinates
on the horizon
sufficiently close to the bifurcation surface, and
transforming to  the Killing coordinate along the flow will
only make them smaller. Therefore we can indeed conclude
that the pullback of $B_a \tilde{\chi}^a$ to a bifurcate
Killing horizon vanishes, provided $B_a$ is regular at the
bifurcation surface. Thus this term will make no
contribution to the black hole entropy on any slice.

Now consider the second term in eq.~(\ref{Qgeneral}). Since the
Killing field is hypersurface orthogonal at the Killing
horizon, we have  $\nabla_a
\tilde{\chi}_b=w_{[a}\tilde{\chi}_{b]}$ for some $w_a$ defined
on the horizon.
 On an arbitrary cross-section we therefore have $\nabla_a
\tilde{\chi}_b= \gamma\hat{\epsilon}_{ab}+
s_{[a}\tilde{\chi}_{b]}$, where $\hat{\epsilon}_{ab}$ is the
binormal to the  cross-section, $\gamma$ is some function,
and $s^a$ is some spacelike vector tangent to the cross
section. Contracting both sides of this equation with
$\tchi^a$
we find that
$\gamma=1$, so we have
\begin{equation}
\nabla_a \tilde{\chi}_b= \hat{\epsilon}_{ab}+
s_{[a}\tilde{\chi}_{b]}.
\label{binormal}
\end{equation}
The same reasoning as in the previous paragraph leads to the
conclusion that the pullback of the covariant
tensor-valued form
$C_{ab}\tilde{\chi}^b$ vanishes on the
horizon.\cite{pullback}
Thus the second term in (\ref{binormal}) does not make any
contribution to the pullback of $\hat{Q}$, and therefore makes no
contribution to the entropy.  The conclusion is that when
pulled back to
any point of the Killing horizon, $\hat{Q}=C_{ab}\hat{\epsilon}^{ab}
=\tilde{Q}$. Therefore in fact the entropy (\ref{S}) is also given by
$2\pi\oint \tilde{Q}$ over
{\it any} cross-section of the horizon.

In the above arguments as in Ref.~\cite{wald1}, one makes
the assumption that the
horizon is a regular bifurcate Killing horizon, and that
all the fields (not just the metric) are regular at the
bifurcation surface. This is a rather strong assumption,
since there certainly exist stationary spacetimes with
Killing horizons that do not possess a regular bifurcation
surface. How do we know that the stationary solution to
which a black hole settles down might not be of this type?
Recent work by Racz and Wald \cite{raczwald} has
something very important to say on this point. Namely, if
the surface gravity is constant and nonvanishing over a patch of
a Killing horizon (and the patch includes a space-like
cross-section of the horizon),
then there exists  a stationary extension of
the spacetime around that patch which extends all the way
back to a regular bifurcation surface.
This result makes it
seem likely that one can dispense with the need to go back
to the bifurcation surface, and instead give a {\it local}
argument for the vanishing of the pullback of the tensors
$B_a\tchi^a$ and
$C_{ab}\tilde{\chi}^b$ to the horizon, provided one assumes
the surface gravity is constant and nonvanishing.
In fact, such an argument exists,
and will be presented elsewhere\cite{prep2}.

The arguments of  Ref. \cite{raczwald} referred only to the
metric, so they do not guarantee that {\it all} of the (matter)
fields can also be extended in a nonsingular way back to the
bifurcation surface, although that may actually be true.
Certainly one knows that all scalars formed from the background
matter fields and their derivatives (as well as the metric and
the curvature) will be nonsingular at the bifurcation surface
since they are constant under the Killing flow.
This is not sufficient however
since a tensor can diverge even if all scalars formed
by contracting it with with tensors from
some particular class are regular.
For scalar matter fields, we have shown that stationarity is
sufficient to enforce the desired regularity at the bifurcation
surface\cite{prep2}. In other cases, regularity may require
that additional conditions be placed on the matter fields (rather
like the assumption of constant surface gravity for the metric).
In the context of the
local argument above, these results are required so that the
vanishing of the pullbacks still holds in the presence of matter
fields.

\subsection{Nonstationary black holes}

The above arguments establish that, provided a stationary
black hole has a regular Killing horizon, there is
no ambiguity in the entropy and it can be evaluated with an
integrand of the same form on any cross-section of the
horizon. In the nonstationary case, there are
three obvious candidate forms for the entropy:
\beqa
S_1&=& 2\pi\oint Q(\xi^a,\nabla^a\xi^b,\cdots)
\nonumber\\
S_2&=& 2\pi\oint \hat{Q}(\xi^a,\nabla^{[a}\xi^{b]})
\nonumber\\
S_3&=&2\pi\oint \tilde{Q}(\heps_{ab})\ .
\label{choices}
\eeqa
The integrand in the first expression is the full potential $Q$ produced
by the Noether charge construction. As indicated, $Q$ may depend on
arbitrarily high order derivatives of the vector field $\xi^a$.
In $S_2$, all of the higher derivatives have been eliminated from
$Q$ via identities
that would hold if $\xi^a$ were a Killing vector,
yielding $\hat{Q}$. Hence the only remaining dependence on the
vector field is on $\xi^a$ and $\nabla^{[a}\xi^{b]}$.
In the last expression $S_3$, the term proportional to $\xi^a$
is dropped, and
$\nabla_{[a}\xi_{b]}$ is replaced by the binormal $\heps_{ab}$
to the particular slice over which the integral is to be evaluated,
yielding $\tilde{Q}$ in the integrand.
As discussed in the previous subsection, all three of
these expressions yield identical results when pulled back to
a bifurcate Killing horizon, with $\xi^a$ equal to the horizon
generating Killing field.

Now if one wishes to define the entropy of a {\it
nonstationary} black hole, it is not so clear what to do.
There is apparently no preferred choice of vector field  out
of which to construct $S_1$ and $S_2$. Further the ambiguities
explained in section \ref{Noether} can no longer be dismissed,
and all of three of the expressions may contain significant
ambiguities. In the absence of a deeper understanding
of black hole entropy, it would seem that there is no
fundamental criterion that might be imposed in defining the
entropy of a nonstationary black hole, other than
it prove convenient for deriving results about the change of
entropy in dynamical processes. Of course,
if an entropy that satisfies the second law can be defined,
that would be a preferred definition.

Actually, for nonstationary perturbations of stationary solutions,
the entropy is well-defined.
Using the Noether charge approach\cite{wald1}, the first law
has been established for variations from a stationary to an
arbitrary, nonstationary solution. In this law, the entropy of the
nonstationary solution is defined as $S_1$, evaluated at the bifurcation
surface of the stationary background. However, it turns out that
in fact one has $\d S_1=\d S_2=\d S_3$
in this case\cite{wald1}.
Moreover, as required by consistency with the first law,
the ambiguities in $Q$ discussed in sect. 2.1 do not affect $S_1$
at the bifurcation surface for these nonstationary perturbations.
(This is obvious for the first and third ambiguities but requires a
short computation for the second one, where $Q$ is modified by
the addition of $\gamma({\cal L}_{\tchi}\phim)$.\cite{wald1})

Wald proposed $S_3$ as the natural candidate for the entropy
of a general nonstationary black hole
since it is a local geometrical
expression\cite{wald1}. Our results of the previous section
show that this proposal at least has the
merit that for a stationary black hole, it gives the correct
entropy on any slice of a Killing horizon. It also seems to
meet the criterion of convenience when compared to the
alternatives, $S_1$ and $S_2$, since no external vector field is
required. Of course, one must still
resolve the inherent ambiguities in $\tilde{Q}$ in some
fashion.

To define $S_1$ or $S_2$ in a dynamical process joining two
(approximately) stationary black hole states, one
must choose some vector field which agrees with the initial and final
horizon-generating Killing fields.
It would seem natural to demand that the vector field also be tangent
to the event horizon generators
through the
intermediate nonstationary interval
(although this could only
be approximately the case, since the original Killing horizon
must be somewhat inside the event horizon).
With such a vector field, one can construct $Q$ and $\hat{Q}$,
and test $S_1$ and $S_2$ as candidates for the black hole entropy.
Having chosen an arbitrary vector field to define $Q$, one still
has $J=dQ$. Thus during the dynamical process, the changes in
$S_1$ from slice to slice will be given by the flux of the
Noether current through the
intervening segment of the
horizon\cite{wald1}, and the total
change in the entropy between the two stationary stages
will be given by the total flux of Noether current.
It would be interesting if one could establish that
($S_2$ or) $S_3$,
with some particular choice of slicing of the horizon,
coincides with $S_1$ for a particular
choice of vector field. In this case, the change
of ($S_2$ or) $S_3$
could also be connected to a flux of Noether current.

\subsection{Explicit entropy expressions}

In this subsection we will find an explicit expression for
the entropy in a wide class of theories.
(We only retain the contributions denoted as $\tilde{Q}$, above.)
As stated above,
ref.~\cite{wald2} presents an inductive algorithm which can
be used  for constructing the Noether charge $Q$. Two useful
facts which arise from this construction, and which will
simplify our calculations below, are:
if the maximum number of derivatives of $\xi^a$ in $J$ is $k$,
then ({\it i})
the maximum number in $Q$ is $k-1$, and ({\it ii}) the
term in $Q$ with $k-1$ derivatives is algebraically
determined by the term in $J$ with $k$ derivatives.

Let us first compute the entropy for a general Lagrangian of
the following functional form:
$\fL=\fL(\phim_m,\nabla_a\phim_m,g_{ab}, R_{abcd})$, that is
involving no more than second derivatives of the spacetime
metric $g_{ab}$, and first derivatives of the matter fields,
denoted by $\phim_m$.
Since ultimately the potential form is pulled back to the horizon,
where the (pulled-back)  term linear in the Killing  field
$\chi^a$ vanishes, we need only determine the part of $Q$
involving at least first derivatives of $\xi^a$. Hence we
only require the part of $J$ with at least {\it second}
derivatives of $\xi^a$. From (\ref{j}), this is the part of
$\theta({\cal L}_\xi \phim)$ with at least second derivatives
of $\xi^a$, and then it follows from (\ref{dL}) that the
latter is given by the part of $\d L$ involving at least
{\it second} derivatives of the field variations. For the
class of Lagrangians, we are considering, such terms can
only arise from variations of the Riemann tensor.

To implement the form notation in computations, it is
convenient to introduce the notation $\eps_{a_1\cdots a_m}$,
which denotes the volume form $\eps_{a_1\cdots a_D}$
regarded as a tensor valued ($D-m$)-form.
Note that there are therefore many notations for the same
tensor field, and we use whichever one is convenient at any
given juncture.
A result that we will make use of is
\begin{equation}
d(W^{a_1\cdots a_m}\epsilon_{a_1\cdots a_m})=
m(\nabla_b W^{a_1\cdots a_{m-1}b})\epsilon_{a_1\cdots
a_{m-1}}.
\label{forms}
\end{equation}

With the $D$-form $\eps$, the Lagrangian form may be written
as $\fL={\sL}\eps$, where
$\sL=\sL(\phim_m,\nabla_a\phim_m,g_{ab}, R_{abcd})$ is a
scalar function. Now varying $L$, one has
\beqa
\d \fL&=& Y^{abcd}(-2\nabla_a\nabla_c\d g_{bd})\eps + \cdots
\nonumber\\
&=&\nabla_a(-2Y^{abcd}\nabla_c\d g_{bd}\,\eps) + \cdots
\label{xxx}
\eeqa
where ``$\cdots$" indicates terms which make no
relevant contributions to
$Q$, and the tensor $Y^{abcd}$ is defined by
\[
Y^{abcd}\equiv \partial \sL/\partial R_{abcd}\ .
\]
Note that the tensor
$Y^{abcd}$ has all symmetries of Riemann tensor ({\it i.e.,}
$Y^{abcd}=Y^{[ab][cd]}$, $Y^{abcd}=Y^{cdab}$ and
$Y^{a[bcd]}=0$; the last will not be used in the following),
and hence the indices on $\nabla_a\nabla_c\d g_{bd}$ need not
be explicitly antisymmetrized.
Comparing eq.'s (\ref{dL}) and (\ref{xxx}), we see
\beq
\theta=-2 Y^{abcd}\nabla_c \d g_{bd}\,\eps_a + \cdots
\label{anotherlabel}
\eeq
and thus the Noether current is given by
\beqa
J&=&-2 Y^{abcd}\nabla_c(\nabla_{b}\xi_{d}+\nabla_{d}\xi_{b})\eps_a
    +\cdots\nonumber\\
&=&-2 Y^{abcd}\nabla_{(c}\nabla_{b)}\xi_{d}\,\eps_a+\cdots
\label{current}
\eeqa
where now ``$\cdots$" refers to
terms with less than second derivatives of $\xi^a$.

At this point one can apply the formulae of \cite{wald2}
to write down the leading term in $Q$, but it is just
as easy to read it off directly from (\ref{current}) above.
To do so, we must massage (\ref{current}) so that the leading
term appears as an exterior derivative, together with terms
that involve less than two derivatives of $\xi^a$. We have
\begin{eqnarray}
J&=&-2 Y^{abcd}\nabla_{(c}\nabla_{b)}\xi_{d}\,\eps_a+\cdots\cr
&=&-2 Y^{abcd}\nabla_{b}\nabla_{c}\xi_{d}\,\eps_a+\cdots\cr
&=&\nabla_{b}(-2 Y^{abcd}\nabla_{c}\xi_{d}\,\eps_a)+\cdots
\nonumber
\end{eqnarray}
and so
\beq
Q(\xi)=- Y^{abcd}\nabla_c \xi_{d}\,\eps_{ab}+\cdots
\label{almost}
\eeq
where we have dropped the terms which vanish for $\xi^a=0$.
The entropy is given by $S=2\pi\oint Q(\tilde{\chi})$,
where the integral
may be evaluated on any cross-section of the horizon.
One may also construct $\tilde{Q}$ by making the replacement
$\nabla_c\tilde{\chi}_d\rightarrow \heps_{cd}$, which then yields
\beqa
S&=&-2\pi\oint Y^{abcd}\heps_{cd}\eps_{ab}\cr
&=&-2\pi\oint Y^{abcd}\heps_{ab}\heps_{cd}\beps.\label{nice}
\eeqa
In the second line
we have introduced $\beps$, the induced volume form on
the horizon cross-section,
and used the identity $\eps_{ab}=\heps_{ab}\beps$,
which holds when $\eps_{ab}$
is pulled back as a $(D-2)$-form to
the horizon cross-section.
(We have not distinguished the various expressions for $S$ here
since they yield identical results on a Killing horizon.)
This rather general result (\ref{nice})
was also derived in ref.~\cite{visser}.\cite{footn2}

It is a simple exercise to show that eq.~(\ref{nice})
reproduces $S=A/(4G)$ for Einstein gravity, as well as the
expression for the entropy of Lovelock black holes derived
by Hamiltonian methods in ref.~\cite{love}. While in these
examples the black hole entropy depends only on the
intrinsic geometry of the horizon, generically
the entropy (\ref{nice})
depends on both the intrinsic and {\it extrinsic} geometry.
It is
a straightforward exercise to extend this result, for
example,  to Lagrangians including first derivatives of the
Riemann tensor, \ie $\sL=\sL(\phim_m,\nabla_a\phim_m,g_{ab},
R_{abcd}, \nabla_eR_{abcd})$. The final result reduces to
\beq
S=-2\pi\oint \left(Y^{abcd}-\nabla_eZ^{e:abcd}\right)
\heps_{ab}\heps_{cd}\beps\label{nice1}
\eeq
where we have introduced the tensor
$Z^{e:abcd}=\partial\sL/\partial \nabla_eR_{abcd}$. In
principle, this result could have been more complicated
because the Noether current now includes third derivatives
of the vector $\xi^a$, but one finds that the
(anti)symmetries of $Z^{e:abcd}$ reduce the order of the
derivatives in all such terms. Recently, Iyer and Wald have produced
a result, more general than eq.~(\ref{nice1}), for Lagrangians containing
arbitrarily high order derivatives of the curvature\cite{wald1}.

Finally, we would like to provide an explicit illustration of the
ambiguities which arise in these expressions for black hole
entropy.
Consider the following interaction involving the metric and a scalar field
\[
\fL_i=[\nabla^a\nabla^b\phis\nabla_a\nabla_b\phis-(\nabla^2\phis)^2
+R_{ab}\nabla^a\phis\nabla^b\phis]\eps\ \ .
\]
Despite the fact that second derivatives of the scalar appear in $\fL_i$,
by the arguments above,
it is not hard to see that those terms make only contributions to $Q$
with no derivatives of $\xi^a$. Hence eq.~(\ref{nice}) may still
be applied, and from $Y_i^{abcd}=-\nabla^{[a}\phis\,g^{b][c}
\nabla^{d]}\phis$, one finds a contribution to the entropy
\beq
S_i=2\pi\oint g^{\perp}_{ab}\nabla^a\phis\nabla^b\phis\ \bar{\epsilon}
\label{thki}
\eeq
where  $g^{\perp}_{ab}$ is the metric in the subspace normal to
the cross-section on which this expression is evaluated.
Now, in fact, $\fL_i$ is a total derivative:
\[
\fL_i=d\alpha_i=
d[(\nabla^a\nabla^b\phis\nabla_b\phis-\nabla^a\phis\nabla^2\phis)\eps_a].
\]
By the arguments of sect.~2.1, this
only produces a contribution $Q_i=i_{\xi} \alpha_i$
to the Noether potential with no derivatives of $\xi^a$,
which yields a vanishing
contribution to the entropy, $S'_i=0$.
At one level, this apparent contradiction is resolved by the observation
that $S_i$ vanishes as well. To see why, note that
$g^{\perp}_{ab}=\tchi_a\beta_b+\beta_a\tchi_a$, where $\beta^a$ is a
vector orthogonal to the cross-section that satisfies $\beta^a\beta_a=0$
and $\beta^a\tchi_a=1$ at the horizon.
The integrand in eq.~(\ref{thki}) evidently
vanishes, since it is proportional to
$\tchi^a\nabla_a\phis={\cal L}_{\tchi}\phis=0$,
so one does have $S_i=0$ when evaluated on a slice of a Killing
horizon.
On a nonstationary horizon, the Lie derivative would not vanish,
and hence one would expect in general that $S_i\ne S'_i= 0$.

The discrepancy
between the forms of $Q_i$ inferred from (\ref{almost}) and from sect. 2.1,
arises because the ambiguity
$\theta_i\rightarrow\theta_i+d\gamma_i$
in the definition of
$\theta_i$ has implicitly entered our calculations.
The discussion in sect.~2.1, by which $\fL_i=d\alpha_i$ would yield
$Q_i=i_{\xi} \alpha_i$, asserts that $\theta({\cal L}_\xi\phim)={\cal L}_\xi
\alpha_i$. This form explicitly disagrees with the result in
eq.~(\ref{anotherlabel})
if one inserts $Y_i^{abcd}$ as given above, and $\delta g_{bd}=
\nabla_{\!b}\,\xi_d+\nabla_{\!d}\,\xi_b$. However, it
is not hard to show that
this latter result can be re-expressed as $d\gamma_i({\cal L}_\xi\phim)$
up to terms that make no contribution to $\tilde{Q}$.
Alternatively, one may consider $Q$ in eq.~(\ref{almost}) directly
\beqa
Q(\xi)&=&-Y_i^{abcd}\,\nabla_c\xi_d\,\eps_{ab}+\ldots\cr
&=&\nabla^{[a}\phis\,g^{b][c}\nabla^{d]}\phis\,\nabla_c\xi_d
\,\eps_{ab}+\ldots\cr
&=&\left(\nabla^a\phis\nabla^b[{\cal L}_\xi\phis]-{1\over2}\nabla^a\phis
\nabla_c\phis[{\cal L}_\xi g]^{cb}\right)\eps_{ab}+\ldots\cr
&=&\gamma_i({\cal L}_\xi\phim)+\ldots
\nonumber
\eeqa
where as usual we drop terms with no derivatives on $\xi^a$.
Thus the contribution to the entropy calculated in eq.~(\ref{thki})
could be eliminated via the ambiguity in the definition of $\theta$.
This explicitly
illustrates that these ambiguities must be resolved in establishing
a unique definition of black hole entropy for nonstationary horizons.

\section{Field redefinitions}

In this section we introduce a new technique, based on field
redefinitions, for  computing black hole entropy.
If a field redefinition can be used to relate the actions which
govern two theories, then the entropies of black holes
in these theories turn out to be
related via the same field redefinition. Hence one can
determine the entropy for a new theory by
using a field redefinition to transform it to a theory for
which the entropy is already known. This
technique is useful because field redefinitions can
introduce (or remove) certain higher curvature interactions
and other higher derivative terms in a gravitational
action.  In general the expression for black hole entropy
(in particular, $S=A/(4G)$ for Einstein gravity) is modified
by such field redefinitions\cite{foot3}.

The field redefinition technique would not be very practical if
it  weren't for the remarkable fact that the leading order
perturbative result is in fact exact, as one can infer  from
the general
form which the entropy density takes in a Noether charge
derivation.  We will illustrate this perturbative procedure
and its justification with an example below.

The validity of the field redefinition technique rests on
the fact that both the asymptotic structure of the spacetime
and the horizon structure are left intact by the field
redefinitions we consider. To understand this point further,
suppose a metric $\bar{g}_{ab}$ is defined by
$\bar{g}_{ab}\equiv g_{ab}+\Delta_{ab}$, where  $g_{ab}$ is
an asymptotically flat black hole metric,  and $\Delta_{ab}$
is a tensor field constructed from $g_{ab}$ and/or other
tensor fields with the property that it vanishes at
infinity. For example, $\Delta_{ab}$ might be a multiple  of
$R_{ab}$, the Ricci tensor of $g_{ab}$. Then, provided the
tensor field $\Delta_{ab}$ falls off fast enough at
infinity, the mass and angular momentum of the spacetime
given by  $\bar{g}_{ab}$ will be the same as that for
$g_{ab}$. Moreover, if $g_{ab}$ is a stationary black hole
spacetime with a bifurcate Killing horizon generated by a
Killing vector $\chi^a$, and if
${\cal L}_{\chi}\Delta_{ab}=0$, then
$\chi^a$ is a Killing field for $\bar{g}_{ab}$ as well,
and $\bar{g}_{ab}$ has the same Killing horizon and surface gravity
as $g_{ab}$.  The condition that $\Delta_{ab}$ be
invariant under the Killing field is satisfied in our
application since $\Delta_{ab}$ will be constructed from the
metric  and matter fields in a stationary solution of some
theory.
The fact that the Killing horizon and surface
gravity are common to both $g_{ab}$ and $\bar{g}_{ab}$
requires further explanation.

First we reiterate that $\chi^a$ is clearly a Killing field
for $\bar{g}_{ab}$, provided it generates a symmetry of both
$g_{ab}$ and any fields entering $\Delta_{ab}$.  The
bifurcation surface $B$ is defined by the metric-independent
equation $\chi^a=0$, so it must coincide for the two
metrics. The $(D-2)$-dimensional surface $B$ is spacelike with
respect to $g_{ab}$ by assumption, and in fact also with respect to
$\bar{g}_{ab}$. To see why, note that the 2-form
$\bar{\nabla}_a \bar{\chi}_b$
(where $\bar{\chi}_b\equiv \bar{g}_{bc}\chi^c$)
is orthogonal to $B$, and is timelike:
$(\bar{\nabla}_a \bar{\chi}_b)(\bar{\nabla}^a \bar{\chi}^b)
=-(\bar{\nabla}_a \chi^b)(\bar{\nabla}_b\chi^a)
=-({\nabla}_a \chi^b)({\nabla}_b\chi^a)
=-2\kappa^2$,
where $\kappa$ is the surface gravity of the Killing horizon
with respect to $g_{ab}$. To obtain the second equality we
evaluated on $B$ where $\chi^a$ vanishes.
This computation shows not only that $B$ is spacelike with respect
to $g_{ab}$, but also that the surface gravity with respect to
$\bar{g}_{ab}$ agrees with that of $g_{ab}$.

Now if a Killing field
vanishes on a spacelike $(D-2)$-surface $B$, then the
null hypersurface generated
by the null geodesic congruences that start out orthogonal
to $B$ is a Killing horizon\cite{Kay}.
In fact, the Killing horizon generated in this fashion for the
metric $\bar{g}_{ab}$ coincides with that for $g_{ab}$.
This follows because, although the light cones defined by $g_{ab}$
and  $\bar{g}_{ab}$ do not in general agree, it so happens
that the Killing field is null with respect to
$\bar{g}_{ab}$ everywhere on the Killing horizon ${\cal H}$
of $g_{ab}$. To see why, note that
$\bar{g}_{ab}\chi^a\chi^b=\Delta_{ab}\chi^a\chi^b$, which
defines a scalar that is constant along the orbits of the
Killing field. As long as $\Delta_{ab}$ is regular at the
bifurcation surface, where $\chi^a$ vanishes,  this scalar
must therefore vanish everywhere on $\cal H$.

Now consider the first law of black hole mechanics
(\ref{first}) in both  theories. We assume that the field
redefinitions leave the asymptotic properties of the black
holes (\eg the mass and angular momentum)
unchanged in transforming between the two theories.
Now in comparing different black holes,
the extensive variations on the right hand
side of the first law are the same in both theories.
Therefore, since the surface gravities are the same,  the
variation of the  entropy must be the same in the two
theories for all variations. Therefore the entropies must
be the same
up to a constant, within any connected set of stationary black hole
solutions.
Therefore, the entropy  of a black hole in the
theory with action $I[g_{ab},\dots]$ is given by the entropy
in the theory with action $\bar{I}[\bar{g}_{ab},\dots]\equiv
I[g_{ab}(\bar{g},\dots)]$, evaluated on the (common) Killing
horizon, and re-expressed in terms of $g_{ab}$ and the  rest
of  the fields.

We now illustrate the procedure with an example.
Consider the theory governed by the action
\beq
I=\int d^Dx\,\sqrt{-g}\,\left[\frac{R}{16\pi G} + L_m +
\lambda(a_1R^{ab}R_{ab}+a_2R^2)\right]
\label{act1}
\eeq
where $L_m$ is a conventional matter Lagrangian (depending
on no more than first derivatives of the matter fields).
Now consider the following field redefinition
\beqa
&&g_{ab}=\bar{g}_{ab}+16\pi G\lambda\left[ a_1 R_{ab}
            -\frac{\bar{g}_{ab}}{D-2}(a_1+2a_2)R\right. \nonumber\\
&&\phantom{g_{ab}=\bar{g}_{ab}+16\pi G\lambda[ a_1 R_{ab}}
             +\left. 8\pi Ga_1 T_{ab} -
           \frac{8\pi G\bar{g}_{ab}}{(D-2)^2}((D-4)a_1-4a_2)T\right]
\label{redef}
\eeqa
where the energy momentum tensor has the usual definition
\[
T^{ab}=\frac{2}{\sqrt{-g}}\frac{\delta\sqrt{-g}L_m}{\delta g_{ab}}
= {2}\frac{\delta L_m}{\delta g_{ab}}+g^{ab}L_m
\]
and $T=g_{ab}T^{ab}$.
In terms of the field $\bar{g}_{ab}$ the action takes the form
\[
\bar{I}=\int d^Dx\,\sqrt{-\bar{g}}\,\left[\frac{\bar{R}}{16\pi G} +
L_m(\bar{g})+
(8\pi G)^2\lambda(a_1T^{ab}T_{ab}+b_2 T^2) + O(\lambda^2)\right]
\]
where $b_2=(4a_2-(D-4)a_1)/(D-2)^2$.\cite{foot4}
The coefficients in the field redefinition (\ref{redef})
were chosen to eliminate the curvature squared interactions,
and any curvature matter couplings such as
$T^{ab}R_{ab}$ arising at order $\lambda$. The
resulting action is not quite a conventional action for
Einstein gravity coupled to matter fields, since the matter
fields have some higher dimension interactions.
As long as $\rL_m$ contains only first order derivatives of
fields, these extra interactions do as well. From the
Noether charge method, we know that such terms do not lead
to modifications of the black hole entropy
(or the mass, angular momentum,
or any parameters characterizing the black holes at
infinity.)

For the theory governed by $\bar{I}(\bar{g})$, the entropy
is simply given by the standard formula to $O(\lambda^2)$
\beq
\bar{S}=\frac{\bar{A}}{4G}+O(\lambda^2)=\frac{1}{4G}\oint_{\Sigma}
d^{D-2}x\,\sqrt{\bar{h}}+O(\lambda^2)
\label{entbar}
\eeq
where $\bar{h}_{ab}$ is the induced metric on a cross section of the
horizon $\Sigma$. Now expressing the entropy in terms of the original metric,
one finds
\beq
S=\frac{1}{4G}\oint_{\Sigma}d^{D-2}x\,\sqrt{h}\left[1+\frac{1}{2}
h^{ab}\,\delta h_{ab}+O(\lambda^2)\right]
\label{entnew}
\eeq
where $\delta h_{ab}$ is the difference
$\bar{h}_{ab}-h_{ab}$.
The intrinsic metric
may be written as $h_{ab}=g_{ab}-\chi_a \beta_b -\beta_a\chi_b$
where $\chi^a$ is the Killing field and $\beta^a$ is vector field
orthogonal $\Sigma$, satisfying $\beta^a\beta_a=0$ and $\beta^a\chi_a=1$
on the horizon. Thus to first order,
$\delta h_{ab}=\delta g_{ab}-\delta g_{ac}\chi^c\,
\beta_b-\chi_a\delta \beta_b-\delta \beta_a\, \chi_b-\beta_a\delta
g_{bc}\chi^c$,
and since $h^{ab}\chi_b =0=h^{ab}\beta_b$, one
has $h^{ab}\,\delta h_{ab}= h^{ab}\,\delta g_{ab}$.
Therefore
\beqa
&&S=\oint_{\Sigma} d^{D-2}x\,\sqrt{h}\left[\frac{1}{4G}+2\pi\lambda
\left( (a_1+2a_2)R-a_1h^{ab}R_{ab}\right.\right.\nonumber\\
&&\phantom{S=\oint_{\Sigma}
d^{D-2}x\,\sqrt{h}\left[\frac{1}{4G}+2\pi\lambda\right.}
\left.\left.+\frac{8\pi G}{D-2}((D-4)a_1-4a_2)T
-8\pi Ga_1h^{ab}T_{ab}\right)+O(\lambda^2)\right]\ \ .
\nonumber
\eeqa
{}From the Noether charge method, we know that the entropy for
the action (\ref{act1})
can be given entirely by metric
expressions, independent of the matter fields.
Using the leading order equations of motion, namely
\[
T_{ab}=\frac{1}{8\pi G}(R_{ab}-\frac{1}{2}g_{ab}R)
       + O(\lambda)\ \ .
\]
the contributions proportional to $T_{ab}$ can be replaced by
curvature quantities. This yields, up to terms of
$O(\lambda^2)$,
\beqa
S&=&\oint_{\Sigma} d^{D-2}x\,\sqrt{h}\left[\frac{1}{4G}+4\pi\lambda
\left((a_1+2a_2)R-a_1h^{ab}R_{ab}\right)\right]\nonumber\\
 &=&\oint_{\Sigma} d^{D-2}x\,\sqrt{h}\left[\frac{1}{4G}+4\pi\lambda
\left(2a_2R+a_1g_{\perp}^{ab}R_{ab}\right)\right]\ \ .
\label{final1}
\eeqa
where $g_{\perp}^{ab}=g^{ab}-h^{ab}=(\chi^a\beta^b+\beta^a\chi^b)$
is the metric in the subspace normal to the horizon.
In making the perturbative expansion, we have consistently
ignored terms of order $\lambda^2$. Recalling the Noether
charge approach once again we see that, since the action
(\ref{act1}) is linear in $\lambda$, the
modifications to the entropy from higher curvature terms in
the original action would only be linear in $\lambda$.
Therefore the leading order result (\ref{final1}) is in fact
the exact black hole entropy for the action (\ref{act1}).
Let us add that eq.~(\ref{final1})
agrees with the Noether charge result in eq.~(\ref{nice}).

Note that if we had {\it not} accounted for the possible
presence of matter fields, the above method would have led
to a modification of the entropy with precisely one-half the
coefficients given in eq.~(\ref{final1}).
These results are not inconsistent however,
since for any
asymptotically flat vacuum solution for the action
(\ref{act1}), one has that $R_{ab}=O(\lambda)$, and so
in evaluating the entropy, one would find simply
$S=A/(4G)+O(\lambda^2)$ with either formula. In general
though, in the presence of matter or {\it other higher
curvature interactions}, eq.~(\ref{final1}) gives the
correct (exact) modification of the entropy induced by the
interactions appearing in the action (\ref{act1}). Also note
that the terms proportional to $T_{ab}$ in the field
redefinition (\ref{redef}) were required to eliminate
interactions such as $R^{ab}T_{ab}$, which would
arise at $O(\lambda)$, and which would
make contributions to
the black hole entropy. In general then,
when using field redefinitions to reduce
an action with higher curvature interactions to a theory for
which the black hole entropy is known,
it is important to include a matter
Lagrangian $\rL_m$, and to ensure that extra matter
interactions arising after the field redefinition make no
contribution to the black hole entropy
or make contributions one can evaluate.
Note also
that field redefinitions of the matter fields are
allowed,
and can prove quite useful (see below).

The form of the higher curvature interactions for which the
black hole entropy can be determined via field redefinitions
is not completely general. However, this approach provides a
simple method to verify results derived via the more
comprehensive methods now available\cite{wald1,visser}. Note
that field redefinitions of matter fields are also possible
and easily show that many matter interactions do not modify
the black hole entropy despite the fact that they involve
higher derivatives. A simple example of such results is
listed in the last line of Table 1. This result is derived
as follows: Beginning with Einstein gravity coupled to a
scalar field with
$\rL=-{1\over2}((\nabla\phis)^2+m^2\phis^2)$, a field
redefinition $\phis\rightarrow\phis+\lambda\nabla^2\phis$ may
be used to show that an interaction $(\nabla^2\phis)^2$
produces a vanishing entropy density. Similarly the field
redefinition $\phis\rightarrow\phis+\lambda\nabla^4 \phis$
shows the entropy is unmodified by a combination of
interactions, $(\nabla^2\phis)^2$ and
$\nabla^2\phis\,\nabla^4\phis$. Having shown that the first of
these doesn't contribute, it must also be true that the
entropy density vanishes for $\nabla^2\phis\,\nabla^4\phis$.
Working iteratively in this way, it is easy to see that an
arbitrary term $\nabla^{2p}\phis\,\nabla^{2q} \phis$ yields no
contribution to the black hole entropy.
One may similarly arrive at the same conclusion via the Noether
charge technique as well.

\section{Discussion}

Several results have been established in this paper.
These are:
\begin{itemize}
\item
Ambiguities in the definition of the Noether charge $Q$
associated with the horizon generating Killing field
(normalized to unit surface gravity)
have no effect when $Q$ is pulled back to the horizon of a stationary
black hole.

\item
The entropy of a stationary black hole can be expressed as
the integral $2\pi \oint Q$ over
{\it any} cross-section of the horizon,
not just the bifurcation surface.

\item
The pullback of $Q$ to {\it any} cross section of the horizon
can be expressed without reference to the Killing field, yielding
the same expression found by Wald at the bifurcation surface.

\item
The Killing horizon and surface gravity of a stationary black
hole metric
are invariant under field redefinitions of the metric of the form
$\bar{g}_{ab}\equiv g_{ab} + \Delta_{ab}$, where $\Delta_{ab}$ is
a tensor field constructed out of stationary fields.

\item
Using the previous result, a new technique has been developed
for evaluating
the black hole entropy in a given theory in terms of that of another
theory related by field redefinitions. Certain
perturbative, first order, results obtained with this method
are shown to be exact.

\item
The entropy has been evaluated explicitly for black holes in
a wide class of theories using both Wald's Noether charge
approach and the field redefinition method developed in this
paper.
In table 1, we have compiled a list of explicit results for certain
sample higher curvature interactions. The first line of the
table gives the result for the Einstein-Hilbert action as a
reference for our notation. The next two lines are simple
extensions of the results in eq.~(\ref{final1}), to include
potential couplings (\ie no derivatives) of a scalar field
to the curvatures. Note that in these cases, the variation
of the entropy in the first law (\ref{first}) includes
variations of both the metric and the scalar field, on the
horizon. In the third line, $F(\phis,R)$ also generalizes
$R^2$ to an arbitrary polynomial in the Ricci scalar
with scalar-field interactions. The
black hole entropy of the latter interactions have also been
verified by Hamiltonian methods\cite{gk}. The third line
provides the generalization of the results of
ref.~\cite{love} to include scalar potential couplings to
Lovelock curvature interactions. All of these results are
encompassed by the general result in eq.~(\ref{nice1}) which
is listed in fifth line.
\end{itemize}

The results of Ref.'s \cite{wald1,visser,SuWald}
and this paper establish that a first  law of black hole
mechanics holds for black holes with bifurcate Killing horizons
in {\it all} generally covariant theories of
gravity, with the entropy being a local geometrical
quantity given by an integral over an arbitrary cross-section
of the black hole horizon. The generality of this result seems
somewhat surprising. General covariance plays a crucial role in
allowing the total energy and angular momentum to be expressible
as surface integrals at infinity. The underlying reason for the local
geometrical nature of the entropy seems less transparent.

It is not clear how strong a restriction it is to include
only black holes with a bifurcate Killing horizon.
The assumption that
a regular bifurcation surface exists
is not physically well-motivated since,
if a black hole forms from collapse, the bifurcation
surface is not even in the physical spacetime, but only in a
virtual extension thereof.
On the other hand,
if the surface gravity is constant and nonvanishing on a
patch of the horizon including a spacelike cross-section,
then the existence of a regular bifurcation surface (perhaps in
an extension of the spacetime) is guaranteed\cite{raczwald}.
Thus the assumption of a bifurcate Killing horizon is
in fact implied by the zeroth law (constancy of the surface gravity).
In general relativity, with matter
satisfying the dominant energy condition, the zeroth law can be
established from the field equations.
The validity of the zeroth law in other theories
remains an open question that clearly deserves more attention, since
the validity of black hole thermodynamics rests on it.

It is worth emphasizing that it was necessary for us to assume that
not just the curvature but {\it all} the physical fields
and their derivatives are regular
at the bifurcation surface. This
condition follows from stationarity for scalar fields\cite{prep2},
but may require additional assumptions for general matter fields.

Originally, the laws of black hole mechanics were a feature of
classical general relativity\cite{barcarhaw}, and their
relation with thermodynamics was only by way of an analogy.
With the discovery that black holes
radiate quantum mechanically  with a temperature
$\kappa/(2\pi)$\cite{radiate},  the interpretation of
these laws (\eg eq.~(\ref{first})) as true thermodynamic relations became
entirely justified. Yet, the deep significance of the  fact that
classical general relativity already ``knew about" Hawking radiation remains
to be discovered. Can any insight into this mystery be gained
by studying the way  classical black hole thermodynamics
generalizes to arbitrary generally covariant gravity theories?

Within the context of classical general relativity,
Hawking's area theorem\cite{killhorz}
implies that the
``entropy"  can never decrease. Whether or not the entropy
in a general theory satisfies such a second law remains
an open question, although some positive results do
exist, and will be discussed in another paper \cite{2ndlaw}.
Here we just remark
that we have shown, via a field redefinition technique, that the
second law holds in a particular class of theories in which the
gravitational Lagrangian is built algebraically out of the Ricci
scalar.

Although the entropy is always a local geometrical quantity
at the horizon, it is in general not just dependent on the
{\it intrinsic} geometry of a horizon cross-section.
General relativity and the Lovelock theories\cite{locx} are exceptional,
in that the entropy is purely intrinsic.
 Should anything be made of
this distinction? To address this question, it would seem to be
necessary to address a more general question: can one understand the
origin of the ``corrections" to the
area-equals-entropy law in more fundamental terms?

In this regard, it is interesting to consider the
various approaches to deriving black hole entropy from statistical
considerations.
One method is to evaluate the entropy using a stationary point
approximation to the formal
path integral for the canonical partition
function\cite{GibbHawk}
or for the density of states\cite{BrowYork} in quantum gravity.
These manipulations yield the same black hole entropy as that
defined by the first law, and this correspondence should
continue to hold for arbitrary gravitational actions.
This interpretation of black hole entropy thus seems to be robust.

Another approach\cite{thzurek}
locates the entropy in the thermal bath of ambient quantum fields
perceived by stationary observers under the stretched horizon.
In this approach, the gravitational field equations play no role.
It is simply argued that changes in this entropy satisfy the
first law of black hole mechanics. Since the geometrically defined
entropy also satisfies the first law, the two must coincide. This
argument works for any gravitational action,
provided the dynamics leads to a stable equilibrium state.
On the other hand, this approach offers no
insight into why the black hole entropy is expressed in a
particular geometric fashion. Moreover, the total entropy of the
bath is infinite.

If this divergence of the entropy is regulated by imposing
a cutoff of some kind, one can obtain a definite
result for the black hole entropy. This is what is done in
various other approaches\cite{cutoff},
closely related to the membrane
viewpoint of \cite{thzurek}.
In those methods, the black hole entropy
is defined essentially by counting quantum field degrees
of freedom either outside or inside the
horizon. These regulated counting approaches
all yield an entropy proportional to
horizon area in units of the cutoff, basically because the dominant
contribution comes from very short wavelength modes near the horizon.
Choosing the cutoff equal to something of order the Planck length,
one recovers the entropy inferred from the first law
(together with the Hawking temperature)
in Einstein gravity.
How, therefore, can these counting approaches
accommodate the corrections to the
area-equals-entropy law that one finds
from higher derivative terms in the action?

In response to this query we can offer the following observations.
First, it seems likely that if one could carry out the
counting more precisely,
one would find corrections to the entropy
of higher order in the cutoff (Planck) length.
Moreover, the occurrence of curvature quantities
in such corrections would not be too
surprising. In a somewhat analogous context, one obtains
divergent curvature dependent terms
in evaluating the Casimir energy of a curved conducting cavity,
where the curvature refers to the geometry of the cavity
boundary\cite{Candelas}.
The presence of a cutoff then renders these terms finite,
and they have
a physically well-defined origin and value.

Since the counting argument introduces only one
(cutoff) scale, the above  plausibility argument seems to fail
when the higher derivative terms in the Lagrangian have
coefficients whose orders of magnitude
are not all set by the same (Planck) scale.
However, it is conceivable that
dependence of the ``counting entropy"
on the adjustable coefficients in the Lagrangian
might arise via the effect they have on the geometry of
the black hole background in which the counting is done.
It will
be interesting to see whether or not this effect on the
background modes has the right form to reproduce the
entropy defined via the classical first law.
If not, then either this counting interpretation of the black hole
entropy is wrong, or it is correct
and it {\it determines} the coefficients in a curvature expansion
of the entropy, not leaving any adjustable freedom in the Lagrangian
of the theory. That is, in a sense, the entropy functional would
determine the theory.
Indeed one can hope, more generally, that the quest
for a statistical understanding of black hole entropy will lead us
not just to a particular low energy effective Lagrangian, but to
a more fundamental theory of gravity and matter.

\vskip 1cm
We would like to acknowledge useful discussions with
V.~Iyer, J.~Simon, M.~Visser, and especially with R.~Wald.
R.C.M.\ was supported by NSERC of Canada, and Fonds FCAR du
Qu\'{e}bec.  T.J.\ and G.K.\ were supported by NSF Grant~PHY91--12240.
Research at the ITP, UCSB was supported by NSF Grant~PHY89--04035.

\newpage

\def\heps{\hat{\epsilon}}
\def\ssc{\scriptscriptstyle}
\def\al{\alpha}
\def\be{\beta}
\def\rL{\widetilde{L}}

\noindent{\bf TABLE 1: Contributions to black hole entropy from %
higher derivative interactions}
$$\vbox{\offinterlineskip
\hrule
\def\tablerule{\noalign{\hrule}}
\halign{&\vrule#&\strut\quad\hfil#&\strut\quad#\hfil\quad&\strut\quad
#\hfil\quad&\strut\quad#\hfil\quad&\vrule#\cr
height 10pt&\omit&\omit&\omit&\omit&\cr
&\ \ \ \ \ \ &(Interaction)/$\sqrt{-g}$&
(Entropy density)/($4\pi\sqrt{h}$)&Derivation${}^\dagger$&\cr
height 10pt&\omit&\omit&\omit&\omit&\cr
\tablerule
height 10pt&\omit&\omit&\omit&\omit&\cr
&1)&$R$&1&\quad&\cr
&\quad&\quad&\quad&\quad&\cr
&2)&$f(\phis)R_{\mu\nu}R^{\mu\nu}$&$f(\phis)(R-h^{\mu\nu}R_{\mu\nu})
=f(\phis)g_\perp^{\mu\nu}R_{\mu\nu}$ %
                                                          &F,N&\cr
&\quad&\quad&\quad&\quad&\cr
&3)&$F(\phis,R)$&$\partial_RF(\phis,R)$&F,H,N&\cr
&\quad&\quad&\quad&\quad&\cr
&4)&$f(\phis)\sL_p(g)\ {}^\ddagger$&$p\,f(\phis)\sL_{p-1}(h)$&H,N&\cr
&\quad&\quad&\quad&\quad&\cr
&5)&$\sL(\phim_m,\nabla_a\phim_m,g_{ab}, R_{abcd}, \nabla_eR_{abcd})$
&$-{1\over2}\left(Y^{abcd}-\nabla_eZ^{e:abcd}\right)
                         \heps_{ab}\heps_{cd}$&N&\cr
&\quad&\quad&\quad&\quad&\cr
&6)&$\nabla^{2p}\phis\ \nabla^{2q}\phis$&$0$&F,N&\cr
&\quad&\quad&\quad&\quad&\cr
height 10pt&\omit&\omit&\omit&\omit&\cr}

\hrule} $$
\vskip .5cm
\noindent{${}^\dagger$} Methods used to derive the entropy density:
\begin{description}
\item [F] $=$ field redefinition method
\item [H] $=$ Hamiltonian method of ref.~\cite{SuWald}
\item [N] $=$ Noether charge method of ref.~\cite{wald1}
\end{description}

\vskip .5ex

\noindent{${}^\ddagger$} $\sL_p(g)={1\over2^p}%
                      \delta^{a_1b_1\cdots a_pb_p}_{c_1d_1\cdots c_pd_p}%
           R_{a_1b_1}{}^{c_1d_1}\cdots R_{a_pb_p}{}^{c_pd_p}(g), \qquad %
          \sL_{\ssc 0}=1$
\vskip 1.ex


\begin{thebibliography}{999999}
\bibitem
{foot2}
In $D$ dimensions, a cross section of the horizon is a ($D$--2)-dimensional
space, and so the surface area $A$ refers to the volume of such a slice.
\bibitem
{footn}
Here ${\cal L}_\xi$
and $i_\xi$ denote the Lie derivative and interior
multiplication by the vector $\xi^a$, respectively. We use the
identity ${\cal L}_\xi=d\,i_\xi+i_\xi d$. We do not consider
the possibility that an additional exact form appears in the variation
of $L$, which would still be compatible with
diffeomorphism invariance of the theory.
In the case of spinor fields and/or
gauge fields, one must  chose a spin frame or gauge in which
the field components have  vanishing Lie derivatives in
order to apply the method as given in \cite{wald1}.
The identity ${\cal L}_\xi=d\,i_\xi+i_\xi d$ will fix our definition
of the Lie derivative of such gauge- or frame-dependent fields.
\bibitem
{pullback}
It should be emphasized
that it is only the {\it pullback} of $C_{ab}\tilde{\chi}^b$
to the horizon that vanishes by this argument. For example,
the covariant Killing field
$\tilde{\chi}_a$ itself does not vanish on the
horizon, although its pullback does vanish.
\bibitem
{foot4}
Note that to order $\lambda^2$, we are free to interchange $g_{ab}$
and $\bar{g}_{ab}$ in any first order terms.
\end{thebibliography}
\end{document}